\documentclass[a4paper,11pt]{elsart}

\usepackage{natbib}

\usepackage{epsfig}

\usepackage{amssymb}

\begin{document}

\begin{frontmatter}



\title{On Constructing the Analytical Solutions for Localizations in a
Slender Cylinder Composed of an Incompressible Hyperelastic
Material}
\author[dai]{Hui-Hui Dai\corauthref{cor}},
\corauth[cor]{Corresponding author. Tel: +852 27888660; fax: +852
27888561.} \ead{mahhdai@cityu.edu.hk}
\author[hao]{Yanhong Hao},
\author[chen]{Zhen Chen}
\address[dai]{Department of Mathematics and
 Liu Bie Ju Centre for Mathematical Sciences, City University
of Hong Kong, 83 TatChee Avenue, Kowloon Tong, Hong Kong}

\address[hao]{Department of Mathematics, City University of Hong Kong,\\
83 TatChee Avenue, Kowloon Tong, Hong Kong}

\address[chen]{Department of Civil and Environmental Engineering, University of Missouri-Columbia, Columbia, MO 65211-2200 USA}

\begin{abstract}
In this paper, we study the localization phenomena in a slender
cylinder composed of an incompressible hyperelastic material
subjected to axial tension. We aim to construct the analytical
solutions based on a three-dimensional setting and use the
analytical results to describe the key features observed in the
experiments by others. Using a novel approach of coupled
series-asymptotic expansions, we derive the normal form equation of
the original governing nonlinear partial differential equations. By
writing the normal form equation into a first-order dynamical system
and with the help of the phase plane, we manage to solve two
boundary-value problems analytically. The explicit solution
expressions (in terms of integrals) are obtained. By analyzing the
solutions, we find that the width of the localization zone depends
on the material parameters but remains almost unchanged for the same
material in the post-peak region. Also, it is found that when the
radius-length ratio is relatively small there is a snap-back
phenomenon. These results are well in agreement with the
experimental observations. Through an energy analysis, we also
deduce the preferred configuration and give a prediction when a
snap-through can happen. Finally, based on the
maximum-energy-distortion theory, an analytical criterion for the
onset of material failure is provided.
\end{abstract}

\begin{keyword}
Localization; Hyperelasticity; Cylinder; Bifurcations of PDE's

\end{keyword}

\end{frontmatter}

\section{Introduction}
The field of fracture mechanics is becoming extremely broad with the
occurrence of unexpected failure of weapons, buildings, bridges,
ships, trains, airplanes, and various machines. There are two
fundamental fracture criterions: the strain energy release rate
(i.e., G Theory) and the stress intensity factor (i.e., K Theory);
see Arthur and Richard (2002). The experimentally determined stress
intensity depends on the specimen size, and the fracture is
accompanied by energy
localization and concentration.\\
Localization is manifested by degradation of material properties
with localized large deformations, and this feature often results in
formation and propagation of macrocracks through engineering
structures. Due to the importance of localization phenomena in
structural safety assessment, much research has been conducted to
resolve experimental, theoretical and computational issues
associated with localization problems, as reviewed by Chen and
Schreyer (1994) and Chen and Sulsky (1995). For hyperelastic
materials, important progress has been made based on the gradient
approach; see Aifantis (1984) and Triantafyllidis and Aifantis
(1986). However, there is still a lack of analytical results for
three-dimensional boundary-value problems. In this paper, hence, we
study the localization in a slender cylinder composed of an
incompressible hyperelastic material subjected to tension, based on
an analytical approach to solve the three-dimensional governing
equations.
 We also intend to provide
mathematical descriptions for some interesting phenomena as observed
in
experiments.\\
Jansen and Shah (1997) conducted careful experiments on concrete
cylinders by using the feedback-control method. From two test
series, the typical stress-displacement behavior for different
height-diameter ratios with normal strength and high strength was
obtained. It appears that the pre-peak segment of the
stress-displacement curves agrees well with the pre-peak part of the
stress-strain curves, but the post-peak segment shows a strong
dependence on the geometric size (i.e., the radius-length ratio).
More specifically, the longer the specimen is, the steeper the
post-peak part of the stress-displacement curves becomes. Also, they
found that the width of the localization zone changes with the
specimen size. In the experiment by Gopalaratnam and Shah (1985), it
was found that the tangent value in the ascending part of the
stress-strain curves seemed to be independent of the specimen size
but in the post-peak part there was a softening region and no unique
stress-strain relation. Schreyer and Chen (1986) studied the
softening phenomena analytically based on a one-dimensional model.
Their results indicate that if the size of the softening zone is
small enough (in a relative sense), the behavior of
displacement-prescribed loading is unstable, and the softening
curves are steeper than those with a larger size of the softening
region. Here, we shall provide the three-dimensional analytical
solutions to capture
all the localized features mentioned above.\\
Another purpose of this research is to provide a method judging the
onset of failure in a slender cylinder subjected to tension. Here,
we use the maximum-distortion-energy theory (the Huber-Hencky-Von
Mises theory; see Riley et al. 2007), which depicts there are two
portions of the strain energy intensity. One is the portion
producing volume change which is ineffective in causing failure by
yielding, and the other is that producing the change of shape which
is completely responsible
for the material failure by yielding.\\
By constructing the analytical solutions for localizations, it is
possible to get the point-wise energy distribution. Then, an
expression for the maximum value of the strain energy can be
obtained. With the Huber-Hencky-Von Mises theory, we can then
establish an analytical criterion for identifying the onset of failure.\\
Mathematically, to deduce the analytical solutions for localizations
in a three-dimensional setting is a very difficult task. One needs
to deal with coupled nonlinear partial differential equations
together with complicated boundary conditions. Further, the
existence of multiple solutions (corresponding to  no unique
stress-strain relation) makes the problem even harder to solve.
Here, the analysis is carried out by a novel method developed
earlier (Dai and Huo, 2002; Dai and Fan, 2004; Dai and Cai, 2006),
which is capable of treating the bifurcations of nonlinear partial
differential equations. Our results yield the analytical forms of
the strain and stress fields, total elongation, the potential energy
distribution and the strain energy distribution, which are
characterized by localization phenomena. In particular, it is found
that once the localization is formed its width does not change with
the further increase of the total elongation, which is in agreement
with the experimental observations. We also provide a description
for the snap-through
phenomenon.\\
The remaining parts of the paper are arranged as follows. In section
2, we formulate the three-dimensional govering equations for the
axisymmetric deformations of a circular cylinder. We
nondimensionalize them in section 3 to identify the key small
variable and key small parameters. Then, in section 4, a novel
method of coupled series-asymptotic expansions is used to derive the
normal form equation of the original system. By the variational
principle, in section 5, we derive the same equation by considering
the energy. In section 6, with the help of the phase plane, we solve
the boundary-value problems for a given external axial force and a
given elongation, respectively. In section 7, through an energy
analysis, we determine the most preferred configurations and give a
description of the snap-through phenomenon. Also, an analytical
criterion for identifying material failure based on the
Huber-Hencky-Von Mises theory is discussed. Finally, concluding
remarks and future tasks are given in section 8.

\section{\large \bf  Three-dimensional governing equations}
We consider the axisymmetric deformations of a slender cylinder
subjected to a static
 axial force at one plane end with the other plane end fixed.
 The radius of the cylinder is $a$ and the length is $l$. We take a cylindrical coordinate
 system, and denote $(R,\Theta,Z)$ and $(r,\theta,z)$ as the coordinates of a
material
 point of the cylinder in the reference and current
 configurations, respectively. The radial and axial
 displacements can be written as
 \begin{equation}
 u(R,Z)=r(R,Z)-R, w(R,Z)=z(R,Z)-Z.
 \end{equation}
 We suppose that the cylinder is composed of an incompressible hyperelastic material, for which the strain energy density $\Phi$ is a function of the first two
invariants $I_1$ and $I_2$ of the left Cauchy-Green strain tensor,
i.e., $\Phi=\Phi(I_1,I_2)$. Moreover the first Piola-Kirchhoff
stress tensor $\mathbf{\Sigma}$ is given by
\begin{equation}
\mathbf{\Sigma}^T=\frac{\partial \Phi}{\partial
\mathbf{F}}-p\mathbf{F}^{-T},\label{2}
\end{equation}
where $\mathbf{F}$ is the deformation gradient and $p$ is the
indeterminate pressure. If the strains are small, it is possible to
expand the first Piola-Kirchhoff stress components in terms of the
strains up to any order. The expressions for the stress components
are very lengthy, and due to the complexity of calculations, we
shall only work up to the third-order material nonlinearity. The
formula containing terms up to the third-order material nonlinearity
has been provided by Fu and Ogden (1999):
\begin{equation}
\begin{array}{lll}\vspace{.5cm}
\Sigma_{ji}
&=\displaystyle&a_{jilk}^1\eta_{kl}+\frac{1}{2}a_{jilknm}^2\eta_{kl}\eta_{mn}
            \displaystyle+\frac{1}{6}a_{jilknmqp}^3\eta_{kl}\eta_{mn}\eta_{pq}\\
            &
            &\displaystyle+p_0(\eta_{ji}-\eta_{jk}\eta_{ki}+\eta_{jk}\eta_{kl}\eta_{li})
            \displaystyle-p^*(\delta_{ji}-\eta_{ji}+\eta_{jk}\eta_{ki})+\textsl{O}(|\eta_{ij}|^4),
\end{array}\label{ji}
\end{equation}
where \\
$\mathbf{\eta}=\mathbf{F}-\mathbf{I}= \left(
\begin{array}{ccc}\vspace{.3cm}
\frac{\partial u}{\partial R} & 0 & \frac{\partial u}{\partial
Z}\\\vspace{.3cm}
0 & \displaystyle\frac{u}{R} & 0\\
\frac{\partial w}{\partial R} & 0 & \frac{\partial w}{\partial Z}
\end{array}
\right),$\\
$p_0$ is the pressure value in Eq. (2) corresponding to zero
strains, $p^*$ is the incremental pressure, and $a_{jilk}^1,\enspace
a_{jilknm}^2,$ and $a_{jilknmqp}^3$ are incremental elastic moduli
defined by
\begin{equation}
\begin{array}{l}
a_{jilk}^1=\frac{\partial^2\Phi}{\partial F_{ij}\partial
F_{kl}}|_{F=I}, \enspace a_{jilknm}^2=\frac{\partial^3\Phi}{\partial
F_{ij}\partial F_{kl}\partial F_{mn}}|_{F=I}, \\
 a_{jilknmqp}^3=\frac{\partial^4\Phi}{\partial F_{ij}\partial
F_{kl}\partial F_{mn}\partial F_{pq}}|_{F=I}.
\end{array}\label{axishu}
\end{equation}
It can be found that
$$p_0=4\Phi_{01}+2\Phi_{10},$$
where $\Phi_{01}$ denotes the first-order partial derivative of
$\Phi$ with respect to the invariant $I_2$ at
$\mathbf{F}=\mathbf{I}$, $\Phi_{10}$ denotes the first-order partial
derivative of $\Phi$ with respect to the invariant $I_1$ at
$\mathbf{F}=\mathbf{I}$. In the following derivations, we shall also
use $\Phi_{ij}$ to denote the i-th order and the j-th order partial
derivative of $\Phi$ with respect to the invariants $I_1$ and $I_2$
at $\mathbf{F}=\mathbf{I}$. All the coefficients in Eq.
(\ref{axishu}) can be
 expressed in terms of $\Phi_{10},\Phi_{20},\Phi_{01},\Phi_{02},$ and $\Phi_{11},$ and here for brevity the expressions are omitted.\\
 The equilibrium equations for a static and axisymmetric problem are
given by
\begin{equation}
\frac{\partial\Sigma_{zZ}}{\partial
Z}+\frac{\partial\Sigma_{rZ}}{\partial
R}+\frac{\Sigma_{rZ}}{R}=0,\label{e1}
\end{equation}
\begin{equation}
\frac{\partial\Sigma_{rR}}{\partial
R}+\frac{\partial\Sigma_{zR}}{\partial
Z}+\frac{\Sigma_{rR}-\Sigma_{\theta\Theta}}{R}=0.\label{e2}
\end{equation}
The incompressibility condition yields that
\begin{equation}
\eta_{ii}=\frac{1}{2}\eta_{mn}\eta_{nm}-\frac{1}{2}\eta_{ii}^2-\det(\eta_{mn}).\label{t1}
\end{equation}
We consider the case where the lateral surface of the cylinder is
traction-free, then the stress components $\Sigma_{rR}$ and
$\Sigma_{rZ}$ should vanish on the lateral surface. Thus, we have
the boundary conditions:
\begin{equation}
\Sigma_{rR}|_{R=a}=0,\enspace \Sigma_{rZ}|_{R=a}=0.\label{b1}
\end{equation}
Eqs. (\ref{e1})--(\ref{t1}) together with Eq. (\ref{ji}) provide
three governing equations for three unknowns $u,w$ and $p^*$. The
former two are very complicated nonlinear partial differential
equations (PDE's) and the boundary conditions (\ref{b1}) are also
complicated nonlinear relations (cf. (\ref{12})--(\ref{16})). To
describe the localization, one needs to study the bifurcation of
this complicated system of nonlinear PDE's. As far as we know, there
is no available mathematical method. Here, we shall adapt a novel
approach involving coupled series-asymptotic expansions to tackle
this bifurcation problem. A similar methodology has been developed
to study nonlinear waves and phase transitions (Dai and Huo, 2002;
Dai and Fan, 2004; Dai and Cai, 2006). First, we shall
nondimensionalize this system to identify the relevant small
variable and small parameters.
\section{\large \bf Non-dimensionalized Equations}
We introduce the dimensionless quantities through the following
scales:
\begin{equation}
s=l^2\tilde{s},\enspace Z=l\tilde{z},\enspace w=h\tilde{w},\enspace
v=\frac{h}{l}\tilde{v},\enspace
\frac{p^*}{\mu}=\frac{h}{l}\tilde{p}^*,\label{10}
\end{equation}
where $l$ is the length of the cylinder, $h$ is a characteristic
axial displacement and $\mu$ is the material shear modulus, with a
transformation being defined by
\begin{equation}
u=vR,\enspace s=R^2.\label{11}
\end{equation}
Substituting Eqs. (\ref{2}), (\ref{10}) and (\ref{11}) into Eqs.
(\ref{e1})--(\ref{t1}), we obtain
\begin{equation}
\begin{array}{l}
 -p^*_z+b_{14}v_z
+4b_{10}w_s+b_2w_{zz}+s(b_{14}v_{sz}+4b_{10}w_{ss})+\cdots=0,
\end{array}\label{12}
\end{equation}
\begin{equation}
-2p^*_s+8b_2v_s+b_{10}v_{zz}+b_{14}w_{sz}+s4b_2v_{ss}+\cdots=0,\label{13}
\end{equation}
\begin{equation}
\begin{array}{l}
2v+w_z+2sv_s+\varepsilon[v^2+2vw_z+2s(vv_s-v_zw_s\\+v_sw_z)]+\varepsilon^2[v^2w_z+2sv(v_sw_z-v_zw_s)]=0,
\end{array}\label{14}
\end{equation}
where $\varepsilon=\displaystyle\frac{h}{l}$ is regarded as a small
parameter. For convenience, we have replaced
$\tilde{s},\tilde{v},\tilde{w},\tilde{z},\tilde{p}^*$ by
$s,v,w,z,p^*$ in the non-dimensionalized equations. Here and
thereafter, a subscript letter is used to represent the
corresponding partial derivative (i.e., $v_z=\frac{\partial
v}{\partial z}$). The full forms of (\ref{12}) and (\ref{13}) are
very lengthy and we do not write out the nonlinear terms explicitly
for
brevity.\\
Substituting (\ref{10}) and (\ref{11}) into the traction-free
boundary conditions (\ref{b1}), we have
\begin{equation}
\begin{array}{l}\vspace{.2cm}
 -p^*+b_{14}v+2b_1w_z+\nu2b_2v_s
+\varepsilon[p^*v+b_{46}v^2+b_{39}vw_z+b_6w_z^2\\\vspace{.2cm}
+\nu(2p^*v_s+b_{15}vv_s+3b_4v_z^2-2b_{11}v_zw_s+12b_4w_s^2+4b_6v_sw_z)\\\vspace{.2cm}
+\nu^24b_5v_s^2]+\varepsilon^2[-p^*v^2+3b_{47}v^3+30b_7v^2w_z+b_{16}vw_z^2+4b_8w_z^3\\\vspace{.2cm}
+\nu(-2p^*(2vv_s+v_zw_s)+b_{22}v^2v_s+7b_9vv_z^2
+b_{23}vv_zw_s\\\vspace{.2cm}
+28b_9vw_s^2+72b_{12}vv_sw_z+2b_{19}v_z^2w_z+2b_8v_zw_sw_z+8b_{19}w_s^2w_z\\\vspace{.2cm}
+b_{18}v_sw_z^2)+\nu^2(-4p^*v_s^2+3b_{31}vv_s^2+4b_{19}v_sv_z^2+8b_{13}v_sv_zw_s\\+16b_{19}v_sw_s^2+b_{20}v_s^2w_z)+\nu^3b_{21}v_s^3]|_{s=\nu}=0,
\end{array}\label{15}
\end{equation}
\begin{equation}
\begin{array}{l}\vspace{.2cm}
 b_{10}(v_z+2w_s)
+\varepsilon[p^*v_z-b_{41}vv_z+2b_{17}vw_s-b_{11}v_zw_z\\\vspace{.2cm}
+12b_4w_sw_z+\nu(-2b_{11}v_sv_z+24b_4v_sw_s)]+\varepsilon^2[-p^*(vv_z+v_zw_z)\\\vspace{.2cm}
+b_{44}v^2v_z+2b_{16}v^2w_s+b_{30}vv_zw_z+b_{32}vw_sw_z+b_{13}v_zw_z^2\\\vspace{.2cm}
+b_{33}w_sw_z^2+\nu(-2p^*v_sv_z+b_{23}vv_sv_z-b_{3}v_z^3+56b_9vv_sw_s\\\vspace{.2cm}
+2b_{11}v_z^2w_s-12b_3v_zw_s^2+16b_5w_s^3+2b_8v_zv_sw_z\\
+16b_{19}v_sw_sw_z)+\nu^2(4b_{13}v_s^2v_z+16b_{14}v_s^2w_s)]|_{s=\nu}=0,
\end{array}\label{16}
\end{equation}
where $\nu=\frac{a^2}{l^2}$ is a small parameter for a slender
cylinder.\\
Then, Eqs. (\ref{12})--(\ref{16}) compose a new system of
complicated nonlinear PDE's with complicated boundary conditions,
which is still very difficult to solve exactly. However, it is
characterized by a small variable $s$ and two small parameters
($\varepsilon$ and $\nu$), which permit us to use expansion methods
to proceed further.\\
\textbf{Remark:} The coefficients $b_1, b_2,\cdots$ in Eqs.
(\ref{12})--(\ref{16}) can be expressed in terms of
$\Phi_{10},\Phi_{20},\Phi_{01},\Phi_{02}$ and $\Phi_{11}$, and for
brevity we omit their expressions.
  \section{\large \bf Coupled Series-Asymptotic Expansions } We note that
  $s$ is also a small variable as $0\leq s\leq\nu$. An
  important feature of the system (\ref{12})--(\ref{16}) is that the unknowns $w,v,$ and $p^*$
  become the functions of the variable $z$, the small variable $s$
  and the small parameters $\varepsilon$ and $\nu$, i.e.,
  \begin{equation}w=w(z,s;\varepsilon,\nu),\enspace
  v=v(z,s;\varepsilon,\nu),\enspace
  p^*=p^*(z,s;\varepsilon,\nu).\label{17}
  \end{equation}
  To go further, we assume that $w,v,p^*$ have the following
  Taylor expansions in the neighborhood of the small variable $s=0$:
  \begin{eqnarray}
  p^*=P_0(z;\varepsilon,\nu)+sP_1(z;\varepsilon,\nu)+s^2P_2(z;\varepsilon,\nu)+\cdots,\label{18}\\
  v=V_0(z;\varepsilon,\nu)+sV_1(z;\varepsilon,\nu)+s^2V_2(z;\varepsilon,\nu)+\cdots,\label{19}\\
  w=W_0(z;\varepsilon,\nu)+sW_1(z;\varepsilon,\nu)+s^2W_2(z;\varepsilon,\nu)+\cdots.\label{20}
   \end{eqnarray}
   Substituting Eqs. (\ref{18})--(\ref{20}) into Eq. (\ref{13}) and equating the
   coefficient of $s^0$ to be zero yields that
   \begin{equation}
   \begin{array}{ll}\vspace{.2cm}
   -2P_1+8b_2V_1+b_{10}V_{0zz}+b_{14}W_{1z}
   +\varepsilon(2P_1V_0+2P_{0z}W_1++2P_0(4V_1+W_{1z})\\\vspace{.2cm}
   +b_{18}W_1^2-8b_{11}V_{0z}W_1+b_{24}V_{0z}^2+8b_6V_1W_{0z}
   +6b_4V_{0zz}W_{0z}+2b_{11}W_{0zz}W_1+6b_4V_{0z}W_{0zz}\\\vspace{.2cm}
   +b_{20}W_{0z}W_{1z}+4b_{15}V_0V_1+b_{17}V_0V_{0zz}+b_{25}V_0W_{1z})
   +\varepsilon^2H_1(V_0,V_1,W_0,W_1,P_0)=0.
   \end{array}\label{21}
   \end{equation}
  Similarly, substituting Eqs. (\ref{18})--(\ref{20}) into Eq. (\ref{12}) and setting the
   coefficients of $s^0$ and $s^1$ to be zero, we obtain
   \begin{equation}
   \begin{array}{ll}\vspace{.2cm}
   -P_{0z}+4b_{10}W_1+b_{14}V_{0z}+b_2W_{0zz}+\varepsilon(P_{0z}W_{0z}
   +P_0(2V_{0z}+W_{0zz})+4b_{17}V_0W_1+b_{26}V_0V_{0z}\\\vspace{.2cm}
   +2b_6V_0W_{0zz}+24b_4W_1W_{0z}+b_{20}V_{0z}W_{0z}
   +2b_{27}W_{0z}W_{0zz})+\varepsilon^2H_2(V_0,V_1,W_0,W_1,P_0)=0,
   \end{array}\label{22}
   \end{equation}
   \begin{equation}
   \begin{array}{ll}\vspace{.2cm}
   -P_{1z}+16b_{10}W_2+2b_{14}V_{1z}+b_2W_{1zz}+\varepsilon(P_{1z}W_{0z}
   +P_1(4V_{0z}+W_{0zz})+P_{0z}W_{1z}\\\vspace{.2cm}
   +P_0(4V_{1z}+W_{1zz})-2b_{11}W_1V_{0zz}+6b_4V_{0z}V_{0zz}+96b_4W_2W_{0z}
   +2b_{20}V_{1z}W_{0z}+72b_4W_1W_{1z})\\\vspace{.2cm}
   +V_1(16b_7W_1-b_{64}V_{0z}+8b_6W_{0zz})-b_{65}V_{0z}W_{1z}+2b_3(W_{0z}W_{1z})_z
   \\
   +V_0(16b_{17}W_2+2b_{26}V_{1z}+4b_6W_{1zz})+\varepsilon^2H_3(V_0,V_1,W_0,W_1,P_0,P_1)=0.
   \end{array}\label{23}
   \end{equation}
   The expressions of $H_1,H_2$ and $H_3$ are very lengthy, whose expressions are
   omitted for brevity. From the incompressibility condition (\ref{14}), the
   vanishing of the coefficients of $s^0$ and $s^1$ leads to the
   following two equations:
   \begin{equation}
   2V_0+W_{0z}+\varepsilon V_0(V_0+2W_{0z})+\varepsilon^2V_0^2W_{0z}=0,\label{24}
   \end{equation}
     \begin{equation}
   \begin{array}{ll}\vspace{.2cm}
  4V_1+W_{1z}+\varepsilon(-2W_1V_{0z}+4V_1(V_0+W_{0z})+2V_0W_{1z})\\
  +\varepsilon^2V_0(-2W_1V_{0z}+4V_1W_{0z}+V_0W_{1z})=0.
   \end{array}\label{25}
   \end{equation}
  Substituting Eqs. (\ref{18})--(\ref{20}) into the traction-free boundary
  conditions (\ref{15}) and (\ref{16}), and neglecting the terms higher than $O(\nu\varepsilon,\varepsilon^2)$, we obtain
  \begin{equation}
   \begin{array}{ll}\vspace{.2cm}
   -P_0+b_{14}V_0+2b_1W_{0z}
   +\nu(-P_1+b_{70}V_1+2b_1W_{1z})\\\vspace{.2cm}
   +\varepsilon(P_0V_0+b_{46}V_0^2+b_{39}V_0W_{0z}+b_6W_{0z}^2)\\
   +\varepsilon^2H_4(V_0,W_0,P_0)+\nu\varepsilon H_5(V_0,V_1,W_0,W_1,P_0,P_1)=0,
   \end{array}\label{26}
   \end{equation}
  \begin{equation}
   \begin{array}{ll}\vspace{.2cm}
   b_{10}(2W_1+V_{0z})+\nu b_{10}(V_{1z}+4W_2)+\varepsilon(P_0V_{0z}+V_0(-b_{41}V_{0z}\\\vspace{.2cm}
   +2b_{17}W_1)+W_{0z}(-b_{11}V_{0z}+12b_4W_1))+\varepsilon^2H_6(V_0,W_0,W_1,P_0)\\
   +\nu\varepsilon H_7(V_0,V_1,W_0,W_1,W_2,P_0,P_1)=0,
   \end{array}\label{27}
   \end{equation}
   where the lengthy expressions for $H_4-H_7$ are omitted for brevity.
   Eqs. (\ref{21})--(\ref{27}) are seven nonlinear ordinary differential
   equations, which are the governing equations for the seven unknowns
   $P_0,P_1,V_0,V_1,W_0,W_1$ and $W_2$. Mathematically, it is still
   very difficult to solve them directly. To go further, we shall use the
   smallness of the parameter $\varepsilon$.
   From Eq. (\ref{24}), we obtain
   \begin{equation}
   W_{0z}=-V_0(2+\varepsilon V_0)/(1+\varepsilon V_0)^2=-2V_0+3V_0^2\varepsilon-4V_0^3\varepsilon^2+\cdots.\label{28}
   \end{equation}
    Using the above equation in Eq. (\ref{25}), we can express $V_1$ in
    terms of $V_0$ and $W_1$, and then from Eqs. (\ref{21}) and (\ref{23}), we
    can express $P_1$ and $W_2$ in terms of $V_0,W_1$ and $P_0$.
    Substituting the expressions for $W_0,V_1,P_1$ and $W_2$ into
    Eq. (\ref{22}), $P_{0z}$ can be expressed in terms of $V_0$ and $W_1$. Then, from
    Eq. (\ref{26}), $P_0$ can be expressed in terms of $V_0$ and $W_1$.
    Finally, from Eqs. (\ref{22}) and (\ref{27}), we obtain
      \begin{equation}
    \begin{array}{ll}\vspace{0.2cm}
    \displaystyle4b_{10}(W_1-V_{0z})+\varepsilon V_0\frac{b_{81}}{2}(5V_{0z}-2W_1)
    +\nu\frac{b_{10}}{2}(V_{0zzz}+W_{1zz})+\varepsilon^2V_0^2(b_{104}V_{0z}-b_{92}W_1)\\
    \displaystyle+\nu\varepsilon(8b_{27}W_1W_{1z}+b_{28}V_{0z}(V_{0zz}+2W_{1z})
         -b_{10}W_1V_{0zz}+b_{29}V_{0}W_{1zz}+\frac{b_{105}}{2}V_0V_{0zzz})=0,\label{29}
    \end{array}
    \end{equation}
    \begin{equation}
    \begin{array}{ll}\vspace{0.2cm}
    \displaystyle b_{10}(2W_1+V_{0z})+\varepsilon V_0(b_{105}V_{0z}+\frac{b_{81}}{2}W_1)
    +\nu\frac{b_{10}}{2}(V_{0zzz}-6W_{1zz})-\varepsilon^2V_0^2(b_{106}V_{0z}+\frac{b_{92}}{2}W_1)\\
    \displaystyle+\nu\varepsilon(-\frac{b_{81}}{8}W_1W_{1z}-b_{30}V_{0z}W_{1z}+\frac{3}{2}b_{10}W_1V_{0zz}
     -b_{31}V_{0z}V_{0zz}-b_{32}V_{0}W_{1zz}+\frac{b_{105}}{8}V_0V_{0zzz})=0.\label{30}
    \end{array}
    \end{equation}
  We note that the above two equations come from the axial equilibrium equation (the coefficient of $s^0$) and
  the zero tangential force at the lateral surface, the two most important relations for
  tension problems in a slender cylinder.\\
  By eliminating $W_{1zz}$ from Eqs. (\ref{29}) and (\ref{30}) and then expressing $W_1$ in terms of
  $V_0$, finally we obtain an equation for the single unknown $V_0$:
  \begin{equation}
  \begin{array}{ll}
  3b_{10}V_{0z}-3\varepsilon
  b_{28}V_0V_{0z}+3\varepsilon^2b_{110}V_0^2V_{0z}-\frac{3}{4}\nu b_{10}V_{0zzz}
  +\nu\varepsilon b_{34}(V_{0z}V_{0zz}+V_0V_{0zzz})=0.
  \end{array}
  \end{equation}
  By further using Eq. (\ref{28}), we obtain the following equation for the axial strain $W_{0z}$:
  \begin{equation}
   \begin{array}{ll}\vspace{0.2cm}
  3b_{10}W_{0zz}-\varepsilon b_{100}W_{0z}W_{0zz}+3\varepsilon^2b_{110}W_{0z}^2W_{0zz}\\
  -\frac{3}{4}\nu b_{10}W_{0zzzz}
  +\nu\varepsilon b_{35}(4W_{0zz}W_{0zzz}+2W_{0z}W_{0zzzz})=0.\label{36}
  \end{array}
  \end{equation}
  Integrating Eq. (\ref{36}) once, we obtain
  \begin{equation}
   3b_{10}W_{0z}-\varepsilon
  \frac{b_{100}}{2}W_{0z}^2+\varepsilon^2b_{110}W_{0z}^3-\frac{3}{4}\nu b_{10}W_{0zzz}
  +\nu\varepsilon b_{35(}W_{0zz}^2+2W_{0z}W_{0zzz})=C,\label{jifen}
   \end{equation}
  where $C$ is an integration constant. To find the physical meaning of $C$, we consider the resultant force $T$
  acting on the material cross section that is planar and perpendicular to the cylinder axis in the reference configuration, and the formula is\\
  \begin{equation}
  T=\int_0^{2\pi}\int_0^a\Sigma_{zZ}RdRd\Theta.\label{T}
  \end{equation}
 After expressing $\Sigma_{zZ}$ in terms of $W_{0z}$ by using the
 results obtained above, the integration can be carried out, and as
 a result we find that
  \begin{equation}
   \begin{array}{ll}\vspace{0.2cm}
  T=8\pi a^2\mu\varepsilon(3b_{10}W_{0z}-\varepsilon \frac{b_{100}}{2}W_{0z}^2+\varepsilon^2b_{110}W_{0z}^3\\
  -\frac{3}{4}\nu b_{10}W_{0zzz}+\nu\varepsilon
  b_{35}(W_{0zz}^2+2W_{0z}W_{0zzz})).\label{Tzhi}
  \end{array}
  \end{equation}
  Comparing Eqs. (\ref{jifen}) and (\ref{Tzhi}), we have $C=\displaystyle\frac{T}{8\pi a^2\mu\varepsilon}$. Thus, we can rewrite Eq. (\ref{Tzhi}) as
  \begin{equation}
   \begin{array}{ll}\vspace{0.2cm}
  3b_{10}\varepsilon W_{0z}-\frac{b_{100}}{2}(\varepsilon W_{0z})^2+b_{110}(\varepsilon W_{0z})^3-\frac{3}{4}\nu\varepsilon b_{10} W_{0zzz}\\
  +\nu\varepsilon^2 b_{35}(W_{0zz}^2+2W_{0z}W_{0zzz})=\frac{T}{8\pi
  a^2\mu}.\label{40}
  \end{array}
  \end{equation}
  If we retain the original dimensional variable and let $V=W_{0Z}=\varepsilon W_{0z},$ we have
  \begin{equation}
  V+D_1V^2+D_2V^3+a^2(-\frac{1}{4}V_{ZZ}+D_3V_Z^2+2D_3VV_{ZZ})=\gamma,\label{41}
  \end{equation}
  where
  \begin{equation}
  D_1=-\frac{b_{100}}{6b_{10}},D_2=\frac{b_{110}}{3b_{10}},D_3=\frac{b_{35}}{3b_{10}},\gamma=\frac{T}{24b_{10}\pi
  a^2\mu}.\label{42}
  \end{equation}
Since Eq. (\ref{41}) is derived from the three-dimensional field
equations, once its solution is found, the three-dimensional strain
and stress fields can also be found. Also, it contains all the
required terms to yield the leading-order behavior of the original
system. Therefore, we refer Eq. (\ref{41}) as the normal form
equation of the system of nonlinear PDE's (\ref{12})--(\ref{14})
together with boundary conditions (\ref{15}) and (\ref{16}) under a
given axial resultant.
\section{The Euler-Lagrange Equation}
It is also possible to deduce the equation for $V=\varepsilon
W_{0z}$ by considering the total potential energy and then using the
variational principle. By using the expansions obtained in section
4, we can express the two principal invariants $I_1$ and $I_2$ in
terms of $W_{0z}$. The results are
\begin{equation}
\begin{array}{l}\vspace{0.3cm}
I_1-3=\displaystyle3\varepsilon^2W_{0z}^2-2\varepsilon^3W_{0z}^3+2\varepsilon^4W_{0z}^4+s[\varepsilon^2(\frac{81}{64}W_{0zz}^2-\frac{15}{8}W_{0z}W_{0zzz})\\\vspace{0.3cm}
\displaystyle+\varepsilon^3(-\frac{3(13\Phi_{01}+22\Phi_{10})}{32(\Phi_{01}+\Phi_{10})}W_{0z}W_{0zz}^2
+\frac{3(20\Phi_{01}+11\Phi_{10})}{8(\Phi_{01}+\Phi_{10})}W_{0z}^2W_{0zzz})],\\\vspace{0.3cm}
I_2-3=\displaystyle3\varepsilon^2W_{0z}^2-4\varepsilon^3W_{0z}^3+5\varepsilon^4W_{0z}^4+s[\varepsilon^2(\frac{81}{64}W_{0zz}^2-\frac{15}{8}W_{0z}W_{0zzz})\\
\displaystyle+\varepsilon^3(-\frac{3(53\Phi_{01}+71\Phi_{10})}{64(\Phi_{01}+\Phi_{10})}W_{0z}W_{0zz}^2
+\frac{3(25\Phi_{01}+16\Phi_{10})}{8(\Phi_{01}+\Phi_{10})}W_{0z}^2W_{0zzz})].\label{43}
\end{array}
\end{equation}
From Eqs. (\ref{43}), we know the first terms in the right-hand
sides of Eqs. (\ref{43}) are second-order nonlinear. To be
consistent with the third-order material nonlinearity of the stress
components, the strain energy $\Phi$ should be expanded up to the
fourth-order nonlinear terms. So, according to the Taylor's
expansion, we have
\begin{equation}
\begin{array}{l}\vspace{0.2cm}
\Phi=\Phi_{10}(I_1-3)+\Phi_{01}(I_2-3)+\frac{1}{2}[(I_1-3)^2\Phi_{20}\\
+2(I_1-3)(I_2-3)\Phi_{11}+(I_2-3)^2\Phi_{02}]+\cdots.\label{44}
\end{array}
\end{equation}
In Eq. (\ref{44}) it is sufficient to keep the second-order terms of
$I_1-3$ and $I_2-3$. Substituting the expressions of $I_1$ and $I_2$
in Eqs. (\ref{43}) into Eq. (\ref{44}), we have
\begin{equation}
\begin{array}{lll}\vspace{.3cm}
\Phi=\displaystyle
\varepsilon^2\mu(6b_{10}W_{0z}^2-\frac{2}{3}\varepsilon
b_{100}W_{0z}^3+\varepsilon^2b_{110}W_{0z}^4)+s\varepsilon^2\mu
[\frac{81}{64}b_{10}W_{0zz}^2\\
-\frac{15}{64}b_{10}W_{0z}W_{0zzz}+\displaystyle\varepsilon(F_1W_{0z}W_{0zz}^2+F_2W_{0z}^2W_{0zzz})].\label{45}
\end{array}
\end{equation}
 The stored energy per unit length is given by
\begin{equation}
\Psi=\int^a_0 \int^{2\pi}_0 \Phi R d R d \Theta.\label{46}
\end{equation}
Substituting Eq. (\ref{45}) into Eq. (\ref{46}) and carrying out the
integration, we obtain the average stored energy over a cross
section:
\begin{equation}
\begin{array}{lll}\vspace{.3cm}
\tilde{\Psi}=\frac{\Psi}{\pi a^2}=\displaystyle
2\varepsilon^2\mu(6b_{10}W_{0z}^2-\frac{2}{3}\varepsilon
b_{100}W_{0z}^3+\varepsilon^2b_{110}W_{0z}^4)+2a^2\varepsilon^2\mu
[\frac{81}{64}b_{10}W_{0zz}^2\\
\displaystyle-\frac{15}{64}b_{10}W_{0z}W_{0zzz}+\varepsilon(F_1W_{0z}W_{0zz}^2
+F_2W_{0z}^2W_{0zzz})].\label{47}
\end{array}
\end{equation}
Letting $V=W_{0Z}=\varepsilon W_{0z},$ we can further rewrite the
above equation as
\begin{equation}
\begin{array}{lll}\vspace{.3cm}
\tilde{\Psi}=E[\frac{1}{2}V^2+\frac{1}{3}D_1
V^3+\frac{1}{4}D_2V^4+a^2(\displaystyle\frac{27}{256}V_Z^2
-\frac{5}{256}VV_{ZZ}+F_1 V V_Z^2+F_2 V^2 V_{ZZ})],\label{48}
\end{array}
\end{equation}
where $E=24\mu b_{10}$ is the Young's modulus, $F_1$ and $F_2$ are
constants related to
material parameters.\\
The total potential energy for a force-controlled problem is given
by
\begin{equation}
\begin{array}{l}
L=\displaystyle\pi a^2(\int^1_0\tilde{\Psi}d Z-E \int^1_0 \gamma V d Z)\\
\displaystyle=\pi a^2E \int^1_0 (-\gamma
V+\frac{1}{2}V^2+\frac{1}{3}D_1
V^3+\frac{1}{4}D_2V^4+a^2(\displaystyle\frac{27}{256}V_Z^2\\
-\frac{5}{256}VV_{ZZ}+F_1 V V_Z^2+F_2 V^2 V_{ZZ})) d Z.\label{49}
\end{array}
\end{equation}
By the variational principle, we have the following Euler-Lagrange
equation:
\begin{equation}
\frac{\partial L}{\partial V}-\frac{d}{d Z}\frac{\partial
L}{\partial V_Z}+\frac{d^2}{d Z^2} \frac{\partial L}{\partial
V_{ZZ}}=0,\label{50}
\end{equation}
which yields that
\begin{equation}
V+D_1V^2+D_2V^3+a^2(D_3V_Z^2-\frac{1}{4}V_{ZZ}+2D_3VV_{ZZ})=\gamma,\label{52}
\end{equation}
which is just Eq. (\ref{41}). This shows that the normal form equation (\ref{41}) obeys the variational principle for energy.\\
If we multiply $V_Z$ to both sides of Eq. (\ref{52}), it can be
integrated once to yield that
\begin{eqnarray}
\frac{1}{2}V^2+\frac{1}{3}D_1 V^3+\frac{1}{4}D_2
V^4-a^2(\frac{1}{8}V_{Z}^2-D_3 V V_Z^2)=\gamma V+K,\label{53}
\end{eqnarray}
where $K$ is an integration constant. \\
In the following section, we shall discuss the solutions for two
boundary-value problems based on Eqs. (\ref{52}) and (\ref{53}), and
reveal their main characteristics.
\section{Solutions for two boundary-value problems }
We rewrite Eq.(\ref{52}) as a first-order system as follows:
 \begin{eqnarray}
 \begin{array}{l}
 V_Z=y,\\
 y_Z=\displaystyle\frac{-\gamma+V+D_1V^2+D_2V^3+a^2D_3y^2}{a^2(\frac{1}{4}-2D_3V)}.\label{56}
 \end{array}
 \end{eqnarray}
Without loss of generality, we take the length $l$ of the cylinder
to be $1$, then $a$ is equivalent to the radius-length ratio. We
suppose that the two plane ends of the cylinder are attached to
rigid bodies. Then we have
  \begin{equation}
 z=0\hspace{.5cm}(or \enspace constant), \hspace*{1cm}at \hspace*{0.2cm}Z=0,
 1,\label{57}
 \end{equation}
 and\begin{equation}
 r=R, \hspace*{2cm}at \hspace*{0.2cm}Z=0,
 1.\label{58}
 \end{equation}
We point out that although Eq. (\ref{52}) is one-dimensional, it is
derived from the three-dimensional governing equations, and as a
result we can also derive the proper boundary conditions by
considering the condition in the other (radial) dimension such as
Eq. (\ref{58}). If one directly introduces a one-dimensional model
(say, using a gradient theory), such an option is not available. So,
this is another advantage of
Eq. (\ref{52}).\\
From Eqs. (\ref{57}) and (\ref{58}), we have
\begin{equation}
 w_R=0\hspace*{.1cm}(i.e., z_R=0), \hspace*{.2cm}and \hspace*{0.2cm}u_R=0\hspace{.1cm}(i.e., r_R=1)\hspace*{.2cm}at \hspace*{0.2cm}Z=0,
 1.\label{59}
 \end{equation}
 Substituting Eq. (\ref{59}) into Eq. (\ref{t1}) and integrating with respect to
 $R$ once, we obtain
\begin{equation}
w_z=0, \hspace*{2cm}at \hspace*{0.2cm}Z=0,1.\label{60}
\end{equation}
Thus, the proper boundary conditions for Eq. (\ref{52}) are
\begin{equation}
V=0, \hspace*{2cm}at \hspace*{0.2cm}Z=0,1.\label{61}
\end{equation}
To solve this boundary-value problem of the first-order system
(\ref{56}) under Eq. (\ref{61}), we shall conduct a phase-plane
analysis with the engineering stress as the bifurcation parameter.
The critical points of this system is given by
\begin{equation}
y=0, \hspace*{0.8cm}and
\hspace*{0.2cm}V+D_1V^2+D_2V^3=\gamma.\label{62}
\end{equation}
 Here we shall consider a class of strain
energy functions
 $\Phi(I_1,I_2)$ such that the $\gamma-V$ plot based on Eq. $(\ref{62})_2$ has one
 peak, and this requires that
 $$
 D_1<0,\enspace D_2>0,\enspace D_1^2>4D_2.
 $$
 The $\gamma-V$ curves corresponding to Eq. $(\ref{62})_2$ are plotted in
 Fig. 1.
 \begin{center}\includegraphics[scale=0.7]{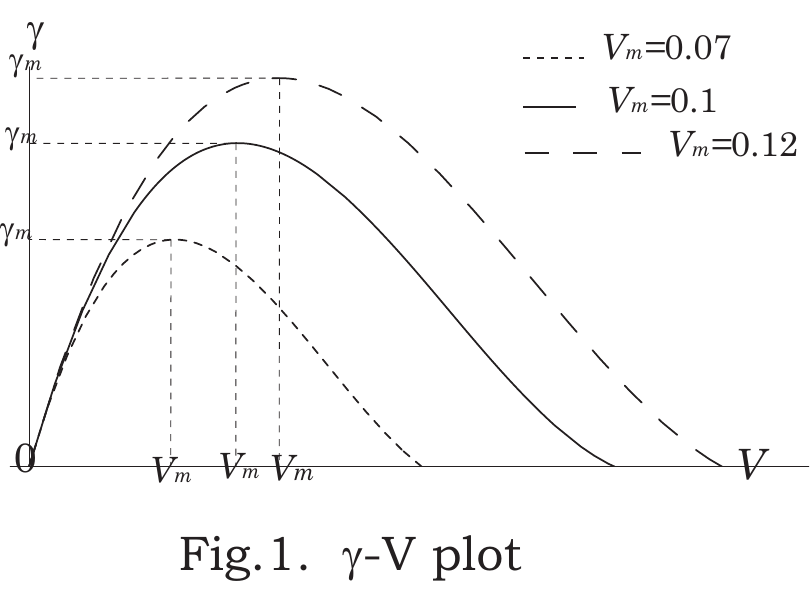}
 \end{center}
 In this figure, $\gamma_m$ is the local maximum of the stress and $V_m$ is the corresponding strain value, and they are given by
  $V_m=-\frac{D_1+\sqrt{D_1^2-3D_2}}{3D_2} ,\enspace \gamma_m=-\frac{2D_1^3+2(D_1^2-3D_2)^{3/2}-9D_1D_2}{27D_2^2}.$
When we take $D_1=-9.45,D_2=22$ and $D_3=-2$, $V_m=0.07$; when we
take $D_1=-6.65,D_2=11$ and $D_3=-2$, $V_m=0.1$; when we take
$D_1=-5.53,D_2=7.6$ and $D_3=-2$, $V_m=0.12$. The three curves in Fig. 1 correspond to these values of $D_1$, $D_2$ and $D_3$, respectively.\\
 In the following discussions we consider the tension case so that
 $\gamma>0$. Similar analysis can be made for the compression case,
 which will not be discussed here. Equation (\ref{53}) can be rewritten as
 \begin{equation}
 V_Z^2=\frac{-K-\gamma
 V+\frac{1}{2}V^2+\frac{1}{3}D_1V^3+\frac{1}{4}D_2V^4}{a^2(\frac{1}{8}-D_3V)}.\label{y59}
 \end{equation}
 In this paper, we consider the case of $D_3\leq 0$. New phenomena can arise for $D_3>0$ and the results will be reported elsewhere.
 For the present case, the phase plane always has a saddle point and a
 center point as $\gamma$ varies, which is shown in Fig. 2.
  \begin{center}\includegraphics[scale=0.6]{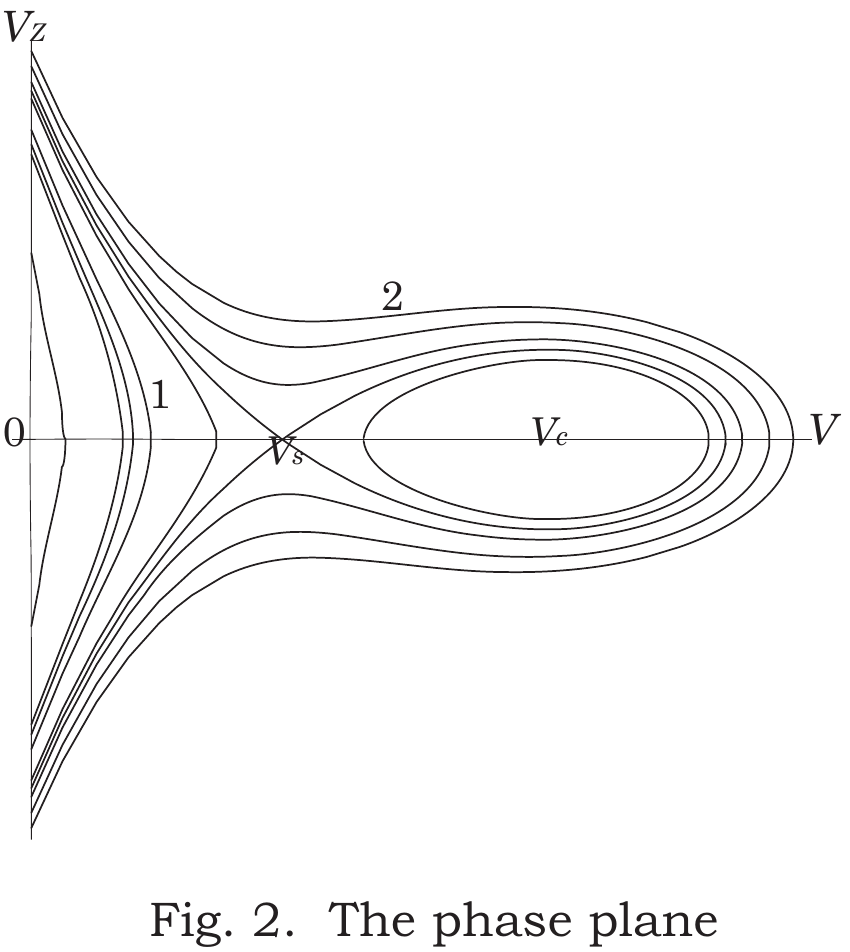}
  \end{center}
   In this figure, $(V_s,0)$ and $(V_c,0)$ are a saddle point and a center
  point, respectively. There are two solutions for the same stress $\gamma$, which are represented by the curve 1 and the curve 2 in
  Fig. 2, respectively. For curve 1, the right hand of Eq. (\ref{y59}) have four real
  roots, which we label in an increasing order by $\alpha_1,g_1,g_2$ and
  $\alpha_2$. We note that the smallest root $\alpha_1$ is smaller than $V_s$. So, from Eq. (\ref{y59})
  we obtain the following expression:
  \begin{equation}
  V_Z=\pm\frac{\sqrt{2D_2}}{a\sqrt{1-8D_3V}}\sqrt{(V-\alpha_1)(V-g_1)(V-g_2)(V-\alpha_2)}.\label{y60}
\end{equation}
Then, an integration leads to
\begin{equation}
 Z=\left\{\begin{array}{l}
 \displaystyle\frac{1}{2}-\frac{a}{\sqrt{2D_2}}\int_V^{\alpha_1}\sqrt{\frac{1-8D_3t}{(t-\alpha_1)(t-g_1)(t-g_2)(t-\alpha_2)}}dt, Z\in[ 0,\frac{1}{2} ]\\
 \displaystyle\frac{1}{2}+\frac{a}{\sqrt{2D_2}}\int_V^{\alpha_1}\sqrt{\frac{1-8D_3t}{(t-\alpha_1)(t-g_1)(t-g_2)(t-\alpha_2)}}dt, Z\in[
 \frac{1}{2},1].
 \end{array}\right.\label{y61}
 \end{equation}
 By Eq. (\ref{61}), $\alpha_1$ can be determined by the following
 two equations:
 \begin{equation}
   \frac{1}{2}=a\sqrt{-D_3}\int_0^{\alpha_1}\sqrt{\frac{t-\frac{1}{8D_3}}{-K-\gamma
   t+\frac{1}{2}t^2+\frac{1}{3}D_1t^3+\frac{1}{4}D_2t^4}},\label{63}
   \end{equation}
   \begin{equation}
 K=-\gamma\alpha_1+\frac{1}{2}\alpha_1^2+\frac{1}{3}D_1\alpha_1^3+\frac{1}{4}D_2\alpha_1^4.\label{y62}
 \end{equation}
  Once $\alpha_1$ is found, the solution corresponding to curve 1 can be obtained from Eq. (\ref{y61}) by numerical integration. In Fig. 3, we have
plotted the solution curves for three different values of the
engineering stress $\gamma$.
    \begin{center}\includegraphics[scale=0.6]{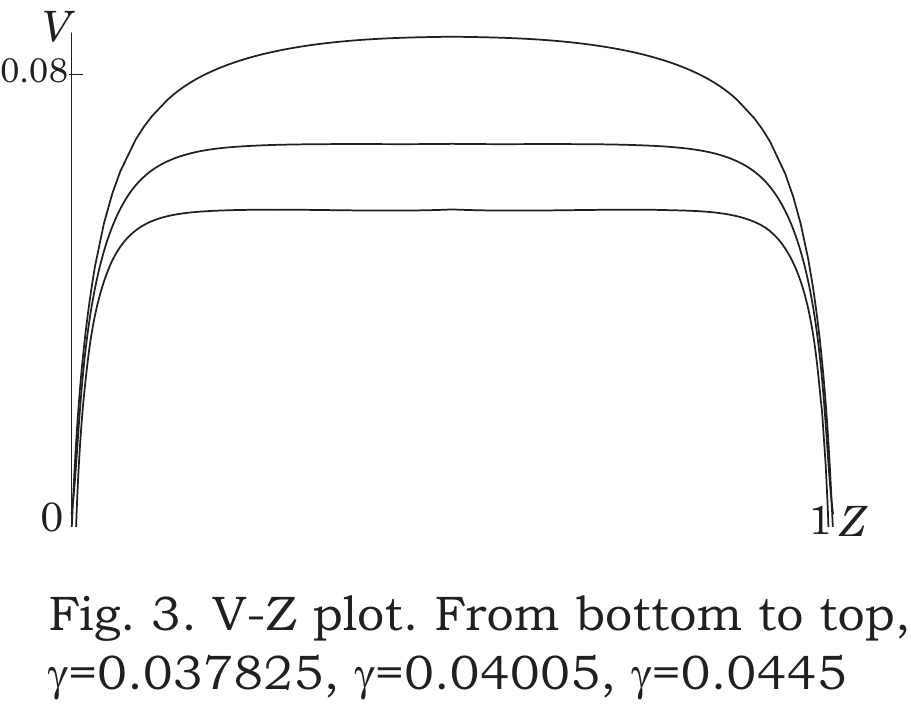}
    \end{center}
     From this figure, we find there is nearly a uniform extension in the middle part, but there are two boundary-layer regions
     near the two ends of the cylinder in order to satisfy the boundary
     conditions.\\
     There is another solution which is represented by curve 2 in
   Fig. 2, and we denote
   the point as $\alpha$ at which $V_Z=0$. Then, Eq. (\ref{y59}) can be rewritten as
  \begin{equation}
  V_Z=\pm\frac{\sqrt{2D_2}}{a\sqrt{1-8D_3V}}\sqrt{(\alpha-V)(\beta-V)[(V-m)^2+n^2]},\label{64}
 \end{equation}
 where $\beta$ is another real root of the right-hand of
 Eq. (\ref{y59}), and $m=-\frac{4D_1+3(\alpha+\beta)D_2}{6D_2}$ and
 $n^2=\displaystyle\frac{-16D_1^2+24(\alpha+\beta)D_1D_2+9D_2(8+(3\alpha^2+2\alpha\beta+3\beta^2)D_2)}{36D_2^2}$.\\
 Then, we obtain
 \begin{equation}
   Z=\left\{\begin{array}{l}\vspace{0.3cm}
  \displaystyle\frac{1}{2}-\frac{a}{\sqrt{2D_2}}\int_V^{\alpha}\sqrt{\frac{1-8D_3t}{(\alpha-t)(\beta-t)[(t-m)^2+n^2]}}dt,Z\in[ 0,\frac{1}{2} ]\\
  \displaystyle\frac{1}{2}+\frac{a}{\sqrt{2D_2}}\int_V^{\alpha}\sqrt{\frac{1-8D_3t}{(\alpha-t)(\beta-t)[(t-m)^2+n^2]}}dt,  Z\in[ \frac{1}{2},1
  ].
    \end{array}\right.\label{65}
 \end{equation}
  By Eq. (\ref{61}), $\alpha$ can be determined by
 \begin{equation}
   \frac{1}{2}=a\sqrt{-D_3}\int_0^{\alpha}\sqrt{\frac{t-\frac{1}{8D_3}}{-K-\gamma
   t+\frac{1}{2}t^2+\frac{1}{3}D_1t^3+\frac{1}{4}D_2t^4}},\label{67}
   \end{equation}
   \begin{equation}
 K=-\gamma\alpha+\frac{1}{2}\alpha^2+\frac{1}{3}D_1\alpha^3+\frac{1}{4}D_2\alpha^4.\label{66}
 \end{equation}
     By numerical integration, we can get $\alpha$ from Eqs. (\ref{66}) and (\ref{67}).
Then the solution corresponding to curve 2 can be obtained from
(\ref{65}) by numerical integration. In Fig. 4, we have plotted the
solution curves for three different values of the engineering stress
$\gamma$.
   \begin{center}\includegraphics[scale=0.6]{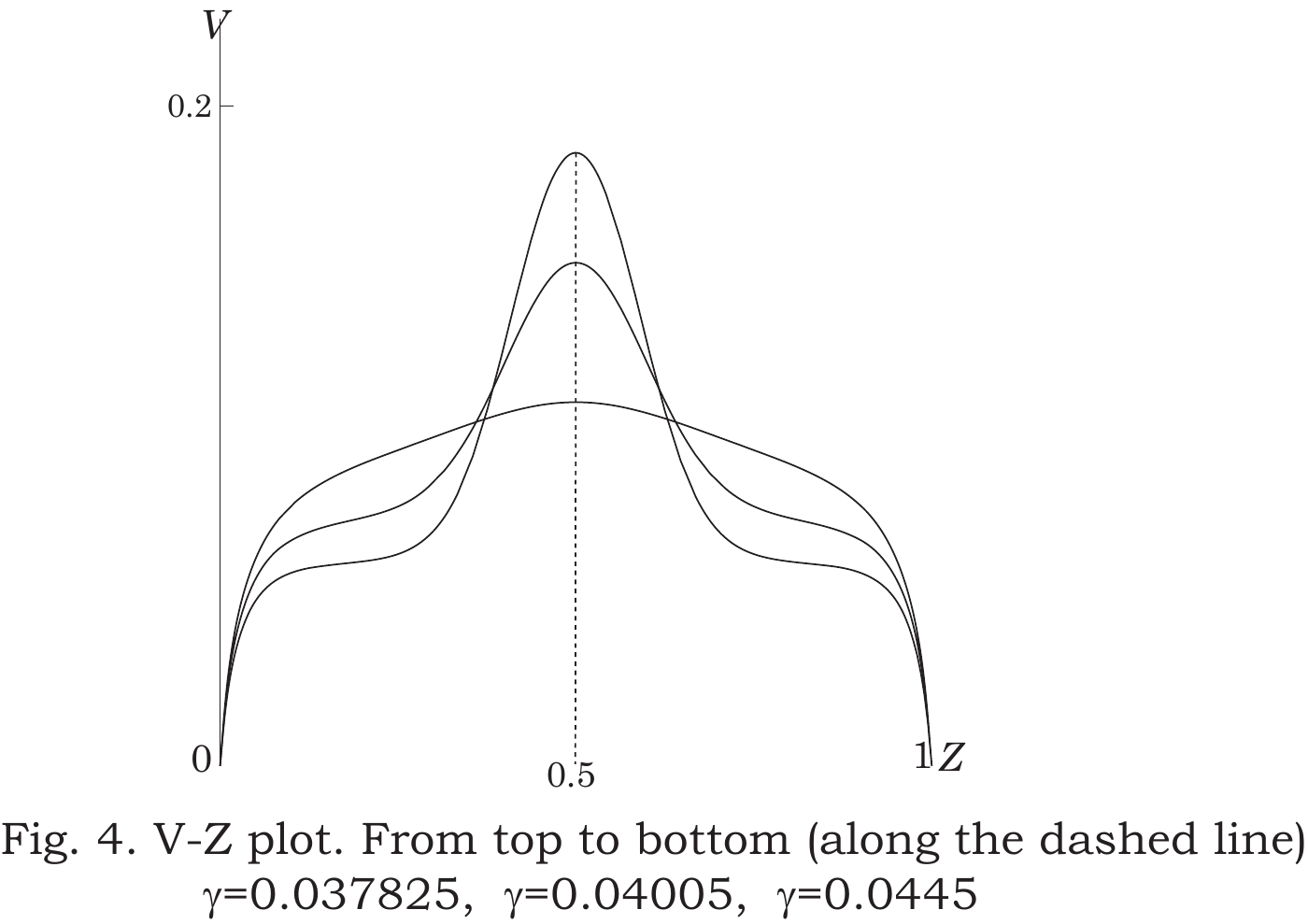}
   \end{center}
   In this figure, there is a sharp-change region in the middle of
   the slender cylinder, that represents the localization and energy concentration. Moreover, the tip is sharper when
   the engineering stress is smaller.\\
   From Eq. (\ref{65}), one can see that the localization solution
   depends on $Z$ through the form $(Z-\frac{1}{2})/a,$ and this implies that the
   localization zone width is proportional to $a$ for a fixed length; see Jansen and Shah (1997).\\
   The solutions obtained above are for a given $\gamma$. To obtain the
   solutions for a displacement-controlled problem, we follow the idea in
   Dai and Bi (2006). For that purpose, we need to get the
   engineering stress-strain curve.\\
   The total elongation is given by
 \begin{equation}
 W_{0}|_{Z=1}-W_0|_{Z=0}=\int_0^1VdZ=\Delta,\label{68}
 \end{equation}
 where the total elongation $\Delta$ is actually the engineering strain since we have taken the length of the cylinder to be
 1. According to the symmetric phase plane and Eqs. (\ref{y61}) and (\ref{65}),
 $V$
 is a function of $Z$, so we can get the total elongation by numerical integrations. In Fig. 5, we have plotted the curves between
 the total elongation $\Delta$ and the engineering stress $\gamma$
 with different material coefficients corresponding to Fig. 1.
    \begin{center}\includegraphics[scale=0.6]{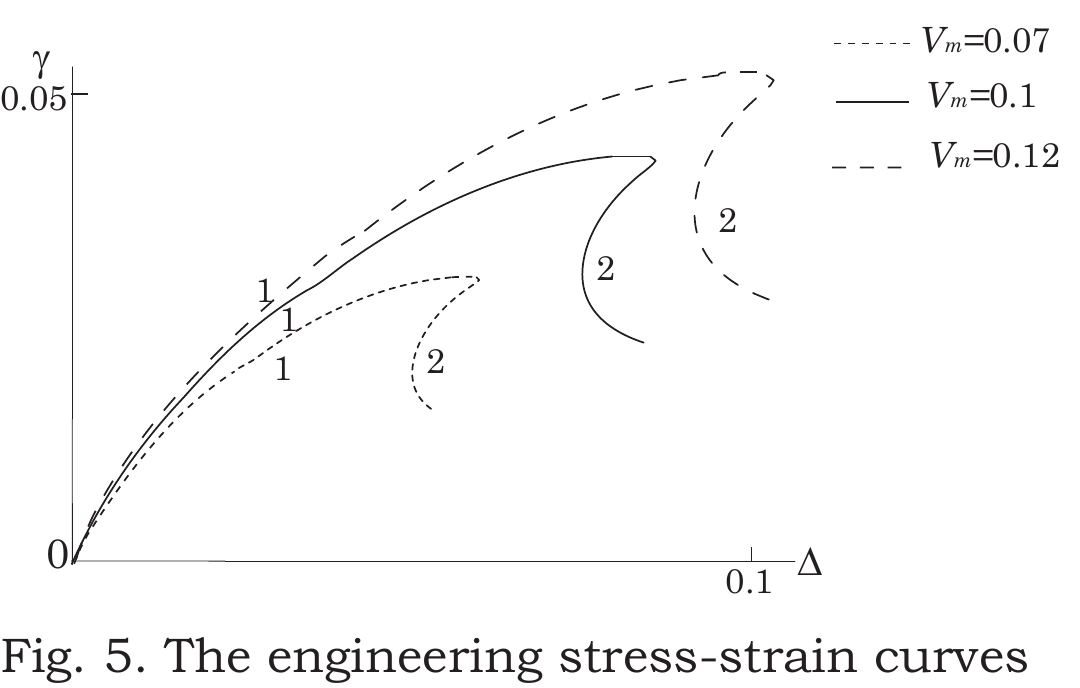}
   \end{center}
  Segment 1 corresponds to the solution given by Eq. (\ref{y61}) (we call it as Solution 1),
  and segment 2 corresponds to the solution given by Eq. (\ref{65}) (we call it as Solution
  2). For a displacement-controlled problem (i.e., given $\Delta$), from Fig. 5, we can get the corresponding $\gamma$ value(s), then the solution(s) is
  given by Eq. (\ref{y61}) or Eq. (\ref{65}).\\
  From Eqs. (1), (\ref{10}), (\ref{11}), (\ref{19}) and (\ref{20}), we can get the shapes
  of the cylinder corresponding to Eq. (\ref{65}) under different material coefficients for a given $\Delta$, which are shown in Fig. 6, where we take
  $D_3=-0.5,F_2=-4,$ and $a=0.04$.
  \begin{center}\includegraphics[scale=0.6]{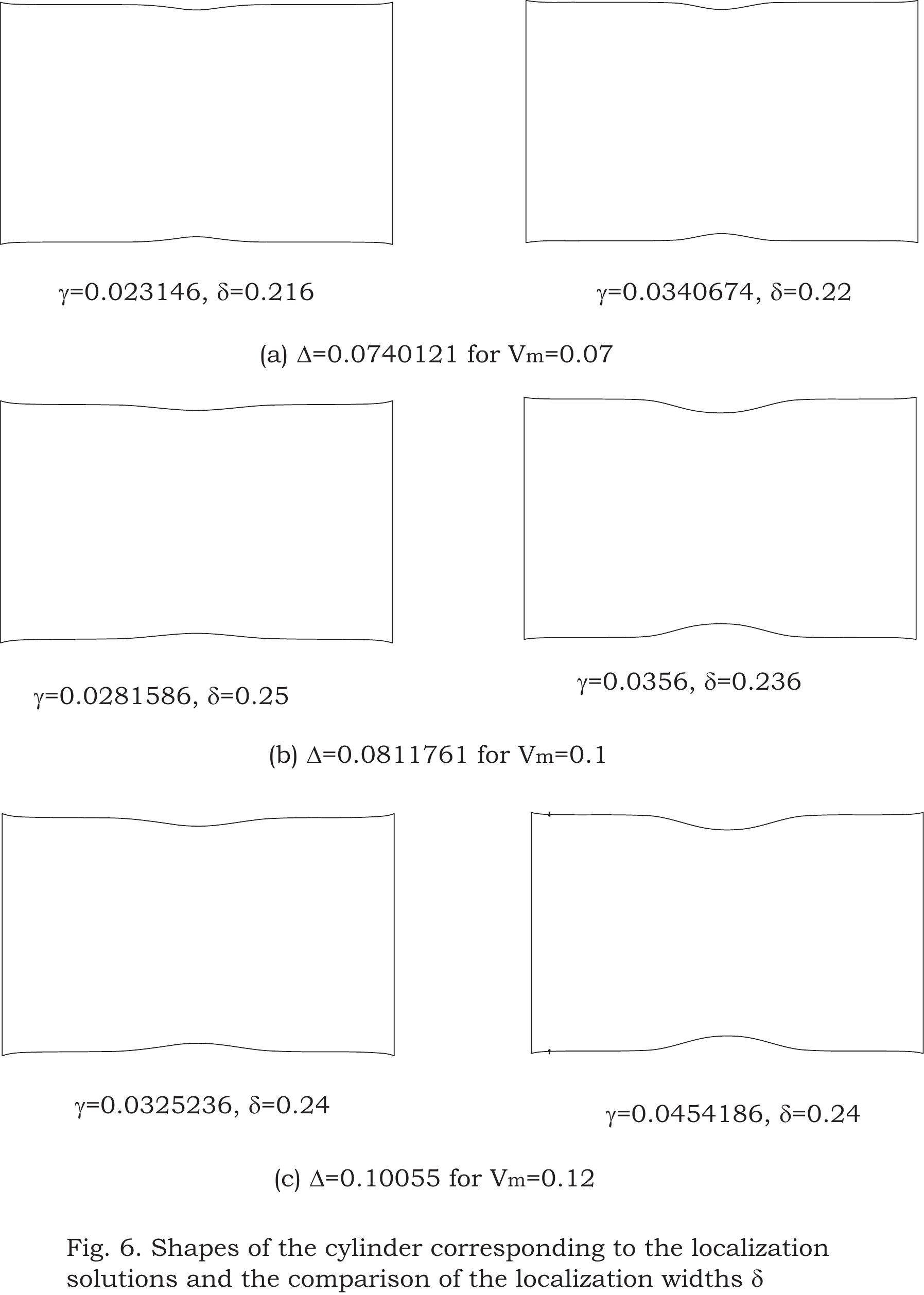}
   \end{center}
 In this figure, the width $\delta$ of the localization is defined as
$1-2\overline{Z}$, where $\overline{Z}$ is the point where the rate
of the slope of the surface radial displacement is the maximum. From
the above figure, one can see that for different material
coefficients the localization widths are different and the
localization width is almost the same for the same material
coefficients with different loads of engineering stress. That is to
say, for different materials the localizations have different
widths, but for the same material, the localization width
is almost uniform during the loading process.\\
Here and thereafter, we fix the material constants to be
$D_1=-6.65,D_2=11,$ and $D_3=-2$. By the same way, we can get the
relations between
 the total elongation $\Delta$ and the engineering stress $\gamma$
 with different radius-length ratios, which are shown in Fig. 7.
\begin{center}\includegraphics[scale=0.6]{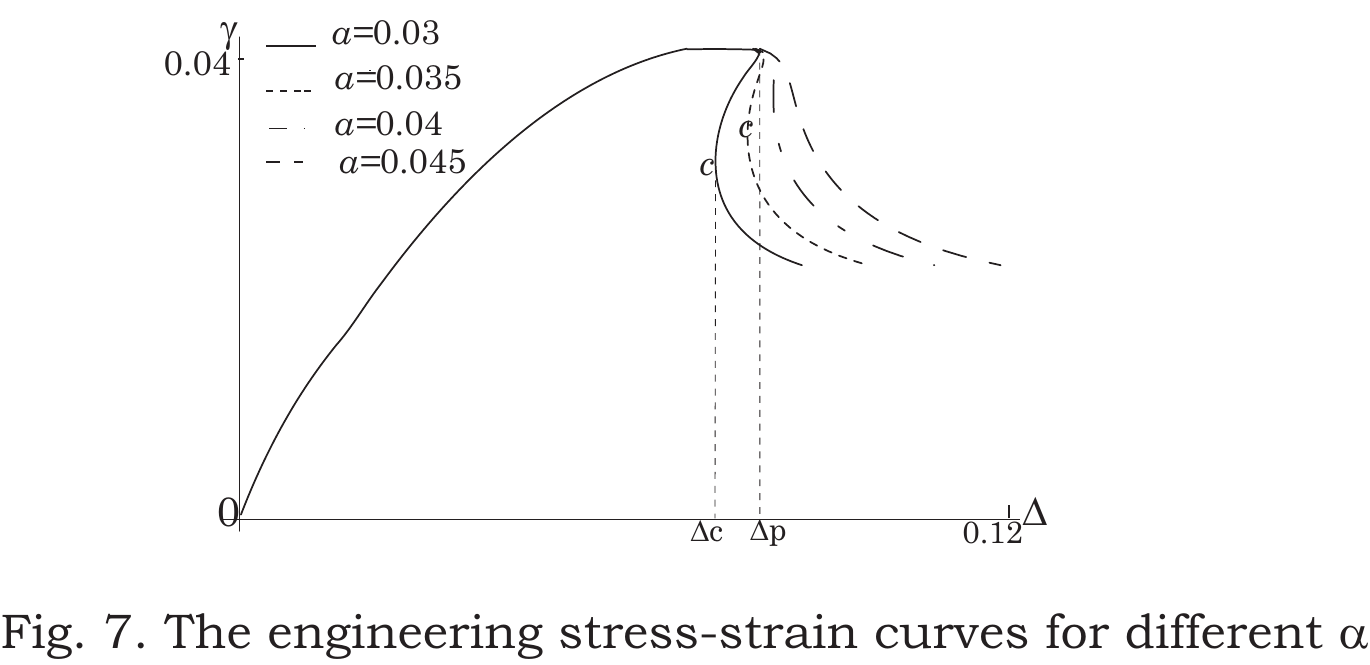}
   \end{center}
  In this figure, we observe that there is a snap-back for a
relatively small value of $a$. We also see that the point $c$
(across which there are multiple values for $\gamma$ for a given
$\Delta$) moves up and toward right as the value of $a$ increases.
For example, when $a=0.03,\enspace \gamma_c=0.0339,$ and when
$a=0.04,\enspace \gamma_c=0.0399$. The post-peak curves show very
significant changes. There is no unique stress-displacement
relationship in the post-peak region. The thinner the specimen is,
the steeper the curve becomes, which is in agreement with the
experimental results by Jansen and Shah (1997). From this figure, we
see that for $a=0.03$, unstable behavior is predicted for a
displacement-controlled loading whereas larger values of $a$ yield
results that are stable. Similar
conclusions follow from the examples given by Schreyer and Chen (1986). \\
As to $a=0.045$, there is a stable relation between the total
elongation $\Delta$ and the engineering stress $\gamma$, which is in
agreement with the experiment result by Gopalaratnam and Shah
(1985), who conducted tensile tests on concrete under carefully
controlled loading conditions and with refined measuring
techniques.\\
 We note that for a displacement-controlled problem, after the elongation $\Delta\geq\Delta_c$ (cf. Fig. 7) there are bifurcations
  from one solution to two solutions (at $\Delta=\Delta_c$), to three solutions ($\Delta_c<\Delta<\Delta_p$), to two solutions
   ($\Delta=\Delta_p$), and to one solution ($\Delta>\Delta_p$). The shapes of the cylinder corresponding to these solutions are
   shown in Fig. 8 for $F_2=-4,$ and $a=0.03$.
   \begin{center}\includegraphics[scale=0.6]{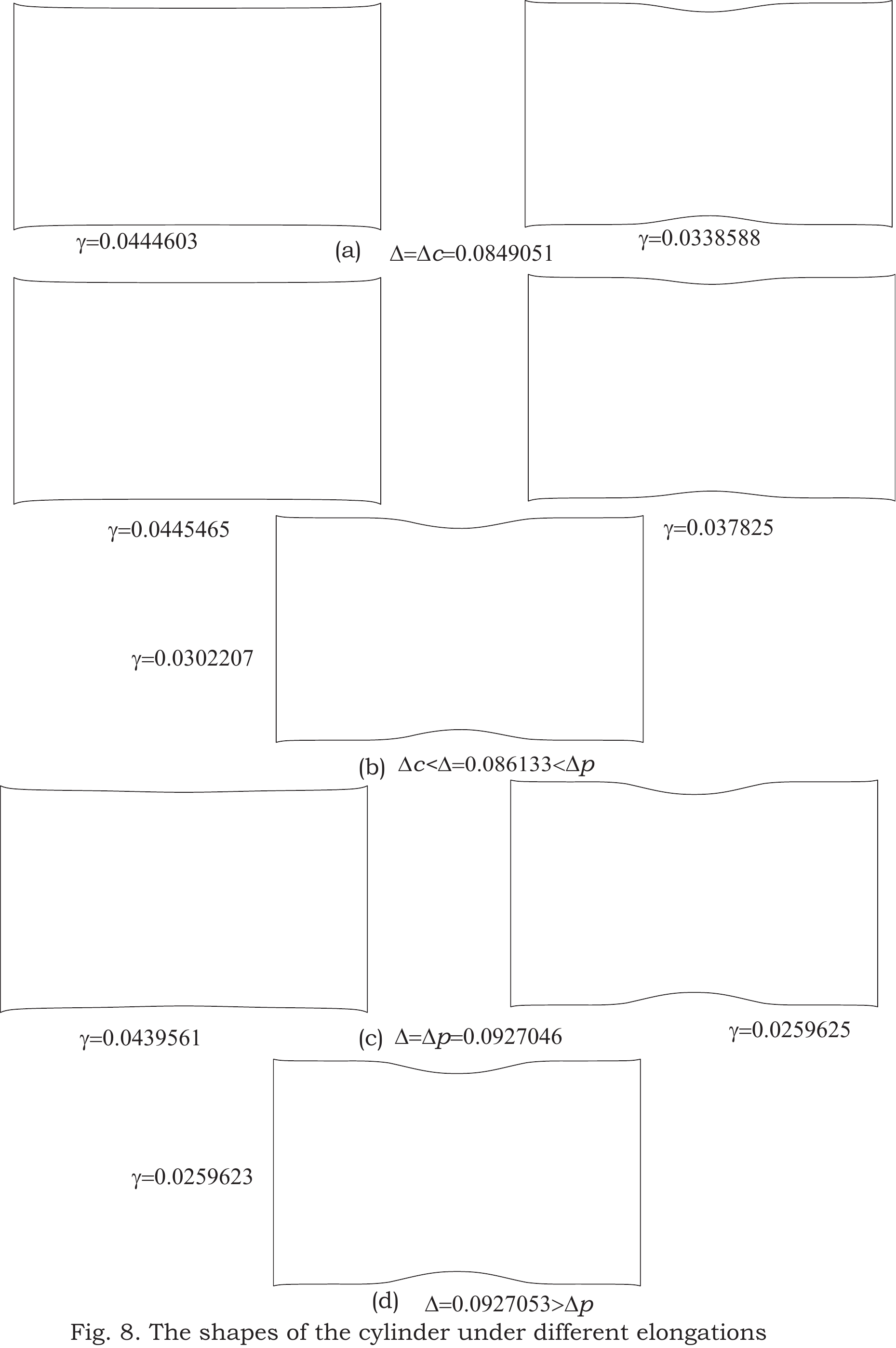}
   \end{center}
 The above figure also manifests that the middle of the cylinder
  becomes
  thinner than the two ends as we pull the slender cylinder.
  The middle part is thinner as the
  engineering stress decreases for a given $\Delta$, which agrees well with the experimental results.
\section{Energy analysis and failure criterion}
As discussed in the previous section, for a relatively small $a$,
there are multiple solutions for $\Delta\geq\Delta_c$. Of course, in
reality only one solution can be observed at one instant. In this
section, we shall further consider the energy values for these
solutions to deduce
which one is most preferred.\\
  From Eq. (\ref{53}), we have
   \begin{equation}V_Z^2=\frac{2(12K+12\gamma V-6V^2-4D_1V^3-3D_2V^4)}{3a^2(-1+8D_3V)}.\label{fang66} \end{equation}
  Substituting Eq. (\ref{fang66}) into Eq. (\ref{52}), we obtain
 \begin{equation}\begin{array}{l}\displaystyle V_{ZZ}=-\frac{4}{3(-1+8D_3V)^2}(3\gamma+24KD_3-3V-3D_1V^2\\-3D_2V^3+12D_3V^2+16D_1D_3V^3+18D_2D_3V^4).\end{array}
 \end{equation}
Then by using Eq. (\ref{49}), we can express the potential energy
(for a given $\gamma$) in terms of $V$ (scaled by $\pi a^2E$):
   \begin{equation}\begin{array}{lll}\vspace{0.2cm}
 P=-\displaystyle\frac{1}{384(-1+8D_3V)^2}(2D_1V^3(-118+15a^2-8192D_3^2V^2\\
 -256(4+3a^2)F_2V+16D_3V(123-5a^2+256(2+a^2)F_2V))\\
 +3(D_2V^4(-59+10a^2-4096D_3^2V^2-512(1+a^2)F_2V\\
 +4D_3V(246-15a^2+256(4+3a^2)F_2V))+2(54K-59V^2\\
 +5a^2V^2+118\gamma V-5a^2\gamma V+4096V^2(K-V(V-2\gamma))D_3^2\\
+256V(4K+V(-(2+a^2)V+(4+a^2)\gamma))F_2\\
+4D_3V(-2(118+5a^2)K+V(246V-5a^2V-492\gamma)\\
+256V(2(-4+a^2)K+V((4+a^2)V-8\gamma))F_2)))).
 \end{array}
   \end{equation}
    The stored energy is given by
\begin{equation}\begin{array}{lll}\vspace{0.2cm}
G=P+\gamma V.
\end{array}
  \end{equation}
  Then from Eqs. (58), (61), (67) and (68), one can calculate the
  energy distributions for a given elongation. The stored energy
  curves corresponding to those values of $\Delta$ in Fig. 8 are
  plotted in Fig. 9. In this figure, labels 1, 2 and 3 correspond to
  different values of $\gamma$ (in a decreasing order). \\
  \begin{center}\includegraphics[scale=0.6]{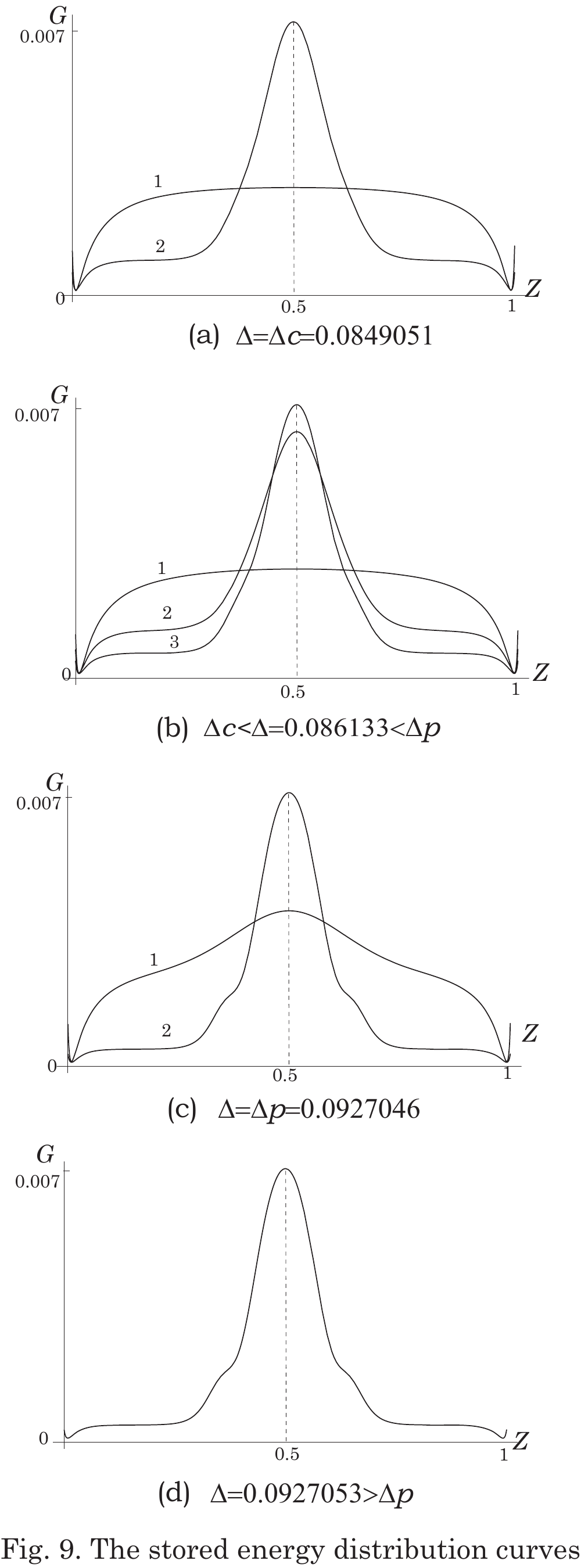}
   \end{center}
  In Fig.9(a), the total stored energy values for curve 1 and
  curve 2 are respectively $G_1^t=0.00241013$ and
  $G_2^t=0.00206265$. Thus for a displacement-controlled problem, as
  $G_2^t<G_1^t$, the shape in the right of Fig. 8(a) represents a
  preferred configuration, and at $\Delta=\Delta_c$ there could be a
  bifurcation from Solution 1 to Solution 2 (a localization
  solution). Correspondingly, there is a snap-through in the
  engineering stress-strain curve at $\Delta=\Delta_c$.\\
  In Fig.9(b), the total stored energy values for curve 1, curve 2
  and curve 3 are respectively $G_1^t=0.00247747,\enspace G_2^t=0.00230487$ and
  $G_3^t=0.00186959$. For a displacement-controlled problem, as
  $G_3^t$ is the smallest, the shape in the bottom of Fig. 8(b)
  represents a preferred configuration.\\
  In Fig.9(c), the total stored energy value for curve 1 and curve
  2 are respectively $G_1^t=0.00275441$ and
  $G_2^t=0.00170114$. For a displacement-controlled problem, as
  $G_2^t<G_1^t$, the shape in the right of Fig. 8(c) represents a
  preferred configuration.\\
  In Fig.10, we have plotted the engineering stress-strain curve
  corresponding to the preferred configuration for a
  displacement-controlled problem. We see that a snap-through takes
  place at $\Delta=\Delta_c$, which leads to the localization (as represented by Solution
  2).
\begin{center}\includegraphics[scale=0.6]{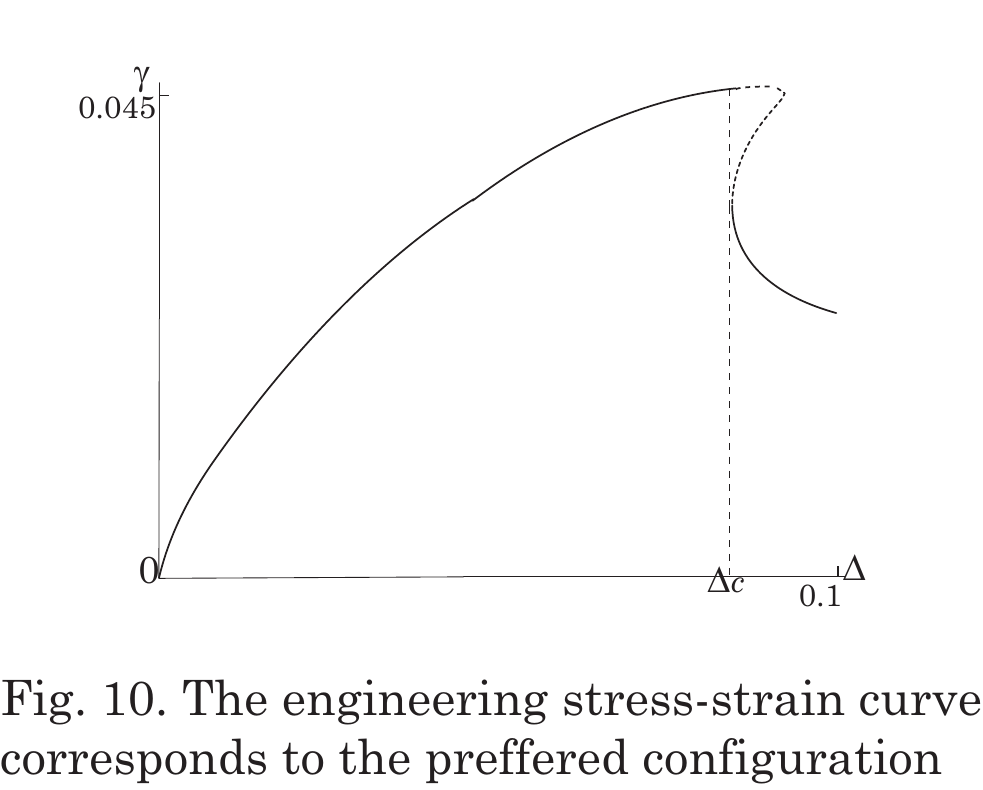}
   \end{center}
   Once the localization happens, there is a high concentration of
  energy around the middle of the cylinder. It is known that if the
  strain energy density reaches a critical value there will be
  the material failure. The analytical results obtained here can be
  used to calculate the stored energy at any material point. The
  largest energy value is attained at $(R,Z)=(a,\frac{1}{2})$ at
  which $V=\alpha$ (cf. Eqs. (61)--(63)). From Eqs. (40), (65), and (66), we
  can express the energy value at this point in terms of $\alpha$,
  and the result is
   \begin{equation}\begin{array}{lll}\vspace{0.2cm}
 G_m=-\frac{1}{384(-1+8D_3\alpha)^2}(2D_1\alpha^3(-118+15a^2-8192D_3^2\alpha^2\\
-256(4+3a^2)F_2\alpha+16D_3\alpha(123-5a^2+256(2+a^2)F_2\alpha))\\
 +3(D_2\alpha^4(-59+10a^2-4096D_3^2\alpha^2-512(1+a^2)F_2\alpha\\
 +4D_3\alpha(246-15a^2+256(4+3a^2)F_2\alpha))+2(54K-59\alpha^2\\
 +5a^2\alpha^2+118\gamma \alpha-5a^2\gamma \alpha+4096\alpha^2(K-\alpha(\alpha-2\gamma))D_3^2\\
 +256\alpha(4K+\alpha(-(2+a^2)\alpha+(4+a^2)\gamma))F_2\\
+4D_3\alpha(-2(118+5a^2)K+\alpha(246\alpha-5a^2\alpha-492\gamma)\\
+256\alpha(2(-4+a^2)K+\alpha((4+a^2)\alpha-8\gamma))F_2))))+\gamma\alpha,
 \end{array}
   \end{equation}
   The values of $G_m$ corresponding to those values of $\Delta$ (in an increasing
   order) in Fig. 8 for preferred configurations are respectively
    $G_m=0.00722628,\enspace G_m=0.00742384,\enspace G_m=0.00742989,$ and $G_m=0.00742989$.\\
Based on the maximum-distortion-energy theory (the Huber-Hencky-Von
Mises theory; see Riley et al. 2007), there are two portions of the
strain energy intensity: one for volume change and the other for
shape change. In the present work, we consider an incompressible
material, so there is no strain energy intensity corresponding to
the volume change. Then the strain energy is only due to distortion.
On the other hand, the strain energy intensity attains its maximum
value at the material point $(a,\frac{1}{2})$. Thus, we can give the
failure criterion
\begin{equation}
 G_m=G_f,
   \end{equation}
   where $G_f$ is the failure value of the strain energy intensity
   for a given material. Fracture will occur
   whenever the energy by Eq. (69) exceeds the limiting value $G_f$.\\

  \section{Concluding Remarks and Future Tasks}

 \hspace{0.5cm} To study the localization phenomenon, a
 phenomenological approach is employed to formulate a
 three-dimensional boundary value problem with an incompressible
 hyperelastic constitutive law. A coupled series-asymptotic
 expansion procedure is developed to solve the non-dimensionalized
 system of governing differential equations with given boundary data
 for a slender cylinder subjected to axial tension. With the
 assumptions appropriate for the slender cylinder, analytical
 solutions have been obtained for the axisymmetric boundary value
 problem, which demonstrate the essential features of localization
 problems and are consistent with the experimental data available.
 Specifically, the width of localization zone depends on the
 material parameters, and it remains unchanged for the same material
 in the post-peak regime. Also, the snap-back and snap-through
 phenomena could be predicted with the analytical approach, and a
 preferred configuration in the post-peak regime could be identified
 via an energy analysis. Due to the lack of three-dimensional
 analytical results available in the open literature, the analytical
 work presented in this paper could complement the analytical,
 experimental and numerical efforts made by the research community
 for the localization problems over the last several decades.\\
 As indicated by Buehler et al. (2003), the hyperelasticity is
 crucial for understanding and predicting the dynamics of brittle
 fracture. Especially, the effect of hyperelasticity is important
 for understanding the failure evolution in nanoscale materials.
 Since localization identifies the onset of material failure, future
 work will focus on the identification of the parameters proposed in
 the current phenomenological model, and on the linkage between the
 continuum and fracture mechanics approaches, via multiscale
 analysis.


 \section*{Acknowledgements}
 The work described in this paper is
supported by two grants from CityU (Project Nos: 7002107 and
7001861).


\end{document}